\documentclass[preprint,twocolumn,tighten]{aastex62}

\usepackage{amsmath,amssymb}
\usepackage{cleveref}
\usepackage{graphicx}
\usepackage{bm}
\usepackage[toc,title,page]{appendix}
\usepackage{tocbibind}
\usepackage{xcolor,ulem}
\usepackage{color}

\definecolor{fgreen}{rgb}{0.13, 0.55, 0.13}

\def\comment#1{}

\usepackage{multirow}

\newcommand{\Msun}{{\rm M}_\odot}

\shorttitle{GW phase shifts in AGN}
\shortauthors{et al.}

\begin{document}
\title{Gravitational Wave Phase shifts of black hole mergers in AGN Disks}

\author{Hiromichi Tagawa}
\email{E-mail: htagawa@shao.ac.jp}
\affil{Shanghai Astronomical Observatory, Shanghai, 200030, People$^{\prime}$s Republic of China}
\author{Connar Rowan}
\affil{Niels Bohr International Academy, The Niels Bohr Institute, Blegdamsvej 17, DK-2100, Copenhagen, Denmark}
\author{János Takátsy}
\affil{Institut für Physik und Astronomie, Universität Potsdam, Haus 28, Karl-Liebknecht-Str. 24-25, Potsdam, Germany}
\author{Lorenz Zwick}
\affil{Niels Bohr International Academy, The Niels Bohr Institute, Blegdamsvej 17, DK-2100, Copenhagen, Denmark}
\affil{Center of Gravity, Niels Bohr Institute, Blegdamsvej 17, 2100 Copenhagen, Denmark}
\author{Kai Hendriks}
\affil{Niels Bohr International Academy, The Niels Bohr Institute, Blegdamsvej 17, DK-2100, Copenhagen, Denmark}
\affil{Center of Gravity, Niels Bohr Institute, Blegdamsvej 17, 2100 Copenhagen, Denmark}
\author{Wen-Biao Han}
\affil{Shanghai Astronomical Observatory, Shanghai, 200030, People$^{\prime}$s Republic of China}
\affil{Hangzhou Institute for Advanced Study, University of Chinese Academy of Sciences, Hangzhou 310124, People’s Republic of China}
\affil{School of Astronomy and Space Science, University of Chinese Academy of Sciences, Beijing 100049, People’s Republic of China}
\affil{Taiji Laboratory for Gravitational Wave Universe (Beijing/Hangzhou), University of Chinese Academy of Sciences, Beijing 100049,
People’s Republic of China}
\affil{Key Laboratory of Radio Astronomy and Technology, Chinese Academy of Sciences, A20 Datun Road, Chaoyang District, Beijing
100101, People’s Republic of China}
\author{Johan Samsing}
\affil{Niels Bohr International Academy, The Niels Bohr Institute, Blegdamsvej 17, DK-2100, Copenhagen, Denmark}
\affil{Center of Gravity, Niels Bohr Institute, Blegdamsvej 17, 2100 Copenhagen, Denmark}

\begin{abstract} 
Ground-based gravitational wave (GW) detectors have discovered about 200 compact object mergers. 
The astrophysical origins of these events are highly debated, and it is possible that at least a fraction of them originate
from dynamical environments. Among these, the disks of active galactic nuclei (AGN) are particularly interesting 
as promising environments, as some observed properties may be more readily produced there. When compact objects merge in these environments, 
acceleration from the central supermassive black hole (SMBH) or nearby companions is inevitable. 
Such acceleration induces a phase shift in the observed GW waveforms, which can serve as a useful tool to distinguish the underlying merging environments for each GW event.
In this paper, we investigate the expected distribution of such acceleration-induced GW phase shifts, using a semi-analytical
model combined with a one-dimensional AGN population synthesis code. 
We find significant contributions from three-body interactions involving a nearby third object. 
Our results indicate that the GW phase shift is likely to be larger compared to other channels,
making it distinguishable by future GW facilities such as TianQin, DECIGO, Taiji, Einstein Telescope, and Cosmic Explorer.
Interestingly, a notable fraction of mergers in fact exhibit a significant GW phase shift ($\gtrsim~{\rm 1\ rad}$) at frequencies
above $10~{\rm Hz}$, which could even be detectable by current GW detectors such as 
LIGO/Virgo/KAGRA. Additionally, if gas-hardening during three-body interactions is taken into account, the GW frequency can
be boosted to $\gtrsim 10~{\rm Hz}$, potentially further aiding in the detection of the phase shift. 
\end{abstract}

\section{Introduction}

Gravitational waves (GWs) from about 200 binary black hole (BBH) mergers 
\citep{LIGO2025_O4aCatalog,LIGO2025_O4aProp}
have been discovered by LIGO/Virgo/Kagra (LVK) \citep{2015CQGra..32g4001L, 2015CQGra..32b4001A, 2021PTEP.2021eA101A}; however, the
underlying formation channels still remain highly debated. Promising channels include 
isolated binary evolution \citep[e.g.][]{Dominik12,Kinugawa14,Belczynski16,Spera19,Tanikawa2022},  
evolution of triple- or quadruple systems
\citep[e.g.][]{Silsbee17,Antonini17,LiuBin2018,Fragione19}, 
dynamical evolution in clusters
\citep[e.g.][]{Banerjee17,Kumamoto18,Rastello18,PortegiesZwart00,Samsing14,OLeary16,Rodriguez16}, 
and active phases of galactic nucleus (AGN) disks
\citep[e.g.][]{Bartos17,Stone17,McKernan17,Tagawa19,Rowan2024_rates,Delfavero2025,Xue2025}.

One of the key observables for distinguishing such formation channels apart is the orbital eccentricity when the GW source
either forms or enters the LVK sensitivity band. Generally, a non-zero eccentricity indicates a dynamical origin,
as something most have perturbed the BBH in question that brought it to merger. Such eccentric GW sources are
expected to form in a wealth of different dynamical
environments \citep[e.g.][]{2018PhRvD..97j3014S, Samsing18, JSDJ18, 2020PhRvD.101l3010S, Antonini2018,Rodriguez2018,Kremer:CMC:2020,Zevin2021,Tagawa20_ecc},
and is therefore an important probe of such systems.
Interestingly, some observed GW sources already hint for a non-zero eccentricity in their GW
form \citep[e.g.][]{2022:Romero-Shaw:GWTC-3-ecc,2024:Gupte:GWTC-3-ecc,Morras2025,deLlucPlanas2025}. A particular interesting event is
GW190521, believed to be associated with either strong precession or high
eccentricity \citep{2020:Romero-Shaw:GW190521,2020:Gayathri:GW190521,2023:Gamba:GW190521,2023:Romero-Shaw:Ecc-or-precc,RomeroShaw2025}, as
well as possessing a relatively large mass \citep{LIGO20_GW190521}, beyond what is expected from pair-instability supernovae \citep{Chatzopoulos12}.

Many dynamical channels will, however, give rise to overlapping eccentricity-, mass- and spin distributions,
and additional measures are therefore needed for disentangling the populations apart. One idea is to look
for perturbations in the GW form, often referred to as GW phase-shift or GW de-phasing, caused by the nearby
environment \citep[e.g.][]{2025ApJ...990..211S, 2024arXiv240804603H, 2025arXiv250324084Z, 2024PhRvD.110j3005Z}.
For example, acceleration from a nearby third BH will result in
a well-defined GW phase-shift \citep{Meiron17,Inayoshi17b,Wong19,Yang2025,Hendriks2024,Zwick2025}, whereas a gaseous
background will cause GW phase shifts through additional energy and momentum-losses during inspiral \citep{ChenRab2025}. In fact, there already exists hints for an accelerated GW source, GW190814 \citep{Yang2025}, which could indicate a merger near a massive object.

The formation of BBH mergers in so-called AGN disks around SMBHs have recently gained significant attention,
as the gas component is a very effective catalyst for capturing the BHs into the disk, even if they did not originally
form there, and bringing them together through migration \citep{Bellovary16,McKernan17,Tagawa19,Grishin2024}.
While such AGN disk models do suggest various unique mass-, spin-,
and eccentricity distributions \citep{Gayathri2021_AGN_O3,Yang19b_PRL,Yang19b,Tagawa20b_spin,Tagawa20_ecc,Tagawa20_MassGap,Vaccaro2024,Xue2025,LiYinJie2025},
similar mergers might form in other environments; however, it remains unexplored if state-of-the-art models for AGN disk mediated
mergers also imply GW mergers with notable GW phase shift, which greatly could help identify their origin using present
and future GW observatories.

With this motivation, we here explore the GW phase shift distribution of BBH mergers forming in AGN disks.
For such mergers, there are at least two sources that can create a notable GW phase shift through acceleration.
The first source is the acceleration from the central SMBH, which all the merging BBHs in the AGN disk experience. 
The second source of acceleration comes from a possible nearby third object if the BBH was driven to merger through a
chaotic interaction with an incoming single BH \citep{Samsing14}, often referred to as a binary-single interaction.
Such chaotic binary-single mergers can occasionally result in exceptionally large phase shifts due to the finite probability
that the third object is very close to the BBH when merging \cite[e.g.][]{Hendriks2024, 2025ApJ...990..211S}. If one takes into account the surrounding gas from the AGN
disk, the characteristic distance between the three objects can greatly decrease during the interaction, resulting in a further significant boost of the GW phase shift \citep[e.g.][]{Rowan2025}. Not only are such high acceleration events expected, but high eccentricity is also anticipated for a significant fraction of AGN mediated mergers \citep{2022Natur.603..237S, 2024MNRAS.535.3630F, 2025arXiv251007952F}.
Thus, GW phase shift and eccentricity can together provide valuable information for identifying BBH mergers in AGN
environments. In addition, modulation is also expected in the GW signal, depending on the type of force and orbital
configurations \citep[e.g.][]{2025ApJ...990..211S, 2024PhRvD.110j3005Z}, which provide possibilities for breaking the
mass-distance degeneracy in the acceleration.

The paper is organized as follow. In Sec. \ref{sec:Methods} we describe our model for how BHs form and interact in AGN disks,
as well as the theory behind GW phase shifts caused by acceleration of nearby objects. Results are presented in Sec. \ref{Sec:Results}, after which we conclude our study in Sec.\ref{sec:conclusions}.

\begin{figure*}
    \centering
    \includegraphics[width=1\linewidth]{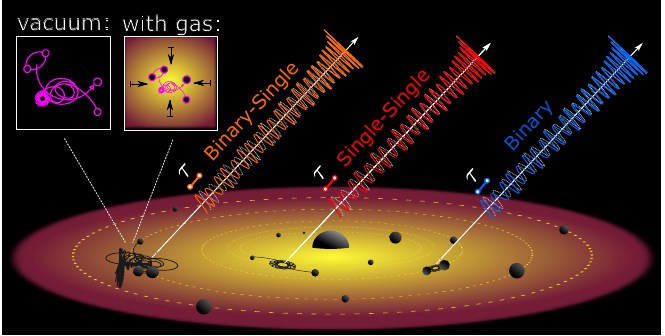}
    \caption{
    A schematic picture of BH mergers within an AGN disk and their GW phase shifts caused by
    acceleration from nearby compact objects.
    Non-GWC binaries and GWC binaries formed through single-single interactions experience strong
    gravitational acceleration due to the central SMBH. A GWC binary formed during binary-single
    interactions undergoes gravitational acceleration from its triple companion.
    In this case, gas drag may significantly harden the triple system,
    thereby boosting the GW frequency.
    }
    \label{fig:schematic}
\end{figure*}

\begin{table*}
\begin{center}
\caption{Model parameters for our fiducial values of model {\bf M1}. In model {\bf M2}, BH interactions occur in a 2D geometry. 
In model {\bf M4}, the initial BH number is reduced to $N_{\rm BH,ini}=6000$ and the initial velocity anisotropy
parameter is enhanced to $\beta_{\rm v}=1$. In model {\bf M12}, migration is not taken into account. 
}
\label{table:parameter_model}
\hspace{-5mm}
\begin{tabular}{c|c}
\hline 
Parameter & Fiducial value \\
\hline\hline
Spacial directions in which BS interactions occur&  isotropic in 3D\\\hline
Mass of the central SMBH & $M_\mathrm{SMBH}=4\times 10^6\,\Msun$ \\\hline
Gas accretion rate at the outer radius of the simulation ($5\,\mathrm{pc}$)
& ${\dot M}_\mathrm{out}=0.1\,{\dot M}_\mathrm{Edd}$ with $\eta=0.1$\\\hline
Fraction of pre-existing binaries & $f_\mathrm{pre}=0.15$ \\\hline
Power-law exponent for the initial density profile for BHs and NSs & $\gamma_\mathrm{\rho,BH}=0$ \\\hline
Initial velocity anisotropy parameter\\such that $\beta_\mathrm{v,BH}v_\mathrm{kep}(r)$ is the BH velocity dispersion  
& $\beta_\mathrm{v,BH}=0.2$ \\\hline
Efficiency of angular momentum transport in the $\alpha$-disk & $\alpha_\mathrm{SS}=0.1$ \\\hline
Stellar mass within 3 pc &$M_\mathrm{star,3pc}=10^7\,\Msun$\\\hline 
Stellar initial mass function slope & $\delta_\mathrm{IMF}=2.35$\\\hline
Angular momentum transfer parameter in the outer star forming regions 
&$m_\mathrm{AM}=0.15$\\
(Eq.~C8 in \citealt{Thompson05}) &\\
\hline
Accretion rate in Eddington units onto\\stellar-mass COs with a radiative efficiency $\eta=0.1$
&$\Gamma_\mathrm{Edd,cir}=1$\\\hline
Numerical time-step parameter &$\eta_t=0.1$\\\hline
Number of radial cells storing physical quantities &$N_\mathrm{cell}=120$\\\hline
Maximum and minimum $r$ for the initial CO distribution&  $r_\mathrm{in,BH}=10^{-6}$ pc, $r_\mathrm{out,BH}=3$ pc \\\hline
Initial number of BHs and NSs within 3 pc &$N_\mathrm{BH,ini}=2\times 10^4$ \\\hline
\label{parameter_model}
\end{tabular}
\end{center}
\end{table*}

\section{Methods}\label{sec:Methods}

\subsection{Overview of models}\label{sec:Overview of models}

We first provide an overview of the processes involved in the evolution and mergers of BHs in an AGN disk. 
We utilize the models from \citet{Tagawa20_MassGap}, where we track $N$-body particles in a 1D-setup that represent BHs as described below.
The BHs are initially part of a nuclear star cluster surrounding the central SMBH, after which they are gradually captured by the
AGN disk through accretion torques and gas dynamical friction \citep{Syer1991,Generozov23,WangY2024_capture,Rowan2025_inclination,Whitehead2025_inclination}.
BHs can also form in the outer regions of the AGN disk, where the gas is gravitationally
unstable \citep{Goodman03,Thompson05,Stone17,Chen2023_SF,EpsteinMartin2025}.
Once embedded in the AGN disk, BHs migrate toward the SMBH due to type I/II torques from the disk \citep{Duffell14,Kanagawa15,Kanagawa18}.
Because migration depends on BH mass, BHs of different masses will encounter each other in the disk, which can lead to binary formation 
primarily due to gas dynamical friction during two-body encounters \citep{Goldreich02,DeLaurentiis2023,Rozner2023,Li2023_BF,Rowan2023,Rowan2024_II,Whitehead2023,Whitehead2025,Qian2024,Dodici2024}
and partially through dynamical interactions during three-body encounters \citep{Tagawa19}. 
After formation, the semi-major axes (SMA) of these binaries generally decrease in the early phases
due to gas dynamical friction \citep{Li2022_I,Li2024_III,Dempsey2022,Dittmann2024,Dittmann2025}.
We also account for subsequent interactions with single stars, other BHs, and other BH binaries. For such few-body scatterings,
we model the resultant changes in the binary SMA, velocities, orbital angular momentum directions, and eccentricities using the prescriptions
outlined in \citet{Leigh18} (see also \citealt{Trani2024}). In the final stage of BBH evolution, 
the separation and eccentricity evolve due to GW emission, with the merger assumed to occur when the pericenter distance becomes smaller than
the innermost stable orbit. After the merger, the remnant BH receives a GW recoil kick following \citet{Lousto12}.
We note that the BH dynamics we track in the AGN disk naturally produce a population of highly eccentric BH binaries,
generated through the GWC mechanism in single-single encounters \citep[e.g.,][]{OLeary09, 2020PhRvD.101l3010S}, and
through binary-single interactions \citep{Samsing14, Samsing18, Tagawa20_ecc, 2022Natur.603..237S}.
Lastly, BHs grow in mass through mergers and Eddington-limited accretion. 

Using this framework, we investigate the evolution of BHs in models~{\bf M1}, {\bf M2}, {\bf M4}, and {\bf M12}, as described in \citet{Tagawa20_ecc}, 
although we do not include neutron stars. 
In short, ~{\bf M1} is our fiducial model, as detailed in \citet{Tagawa20_MassGap}. 
Table~\ref{parameter_model} lists its initial conditions. 
In model~{\bf M2}, binary-single interactions are constrained to a two-dimensional (2D) disk-like geometry instead of the
3D geometry used in ~M1. In model~{\bf M4}, the initial number of BHs is reduced by a factor of $10/3$, and their initial
velocity dispersion is increased by a factor of 5 relative to the fiducial model ~{\bf M1}, 
which assumes a net orbital rotation of the massive objects considering vector resonant relaxation \citep{Szolgyen18}. 
In model ~{\bf M12}, BHs do not migrate within the AGN disk.
We consider these models because the evolution and mergers of BHs in AGN disks are primarily influenced by
the differences they encompass. Compared to the results in \citet{Tagawa20_ecc}, we adjusted the inner radius
within which BHs are removed from $10^{-4}~{\rm pc}$ to $10^{-6}~{\rm pc}$
to ensure that we do not miss mergers occurring near the SMBH. 

Below we outline the theory and calculation of the GW phase shift for the BBH mergers identified in our simulations.

\subsection{Gravitational Wave Phase Shifts}\label{Gravitational Wave Phase Shifts}

When a BBH moves along an accelerated path, its position will deviate from its
unperturbed linear motion by $l = (1/2) a t^2$, where $a$ is the acceleration,
and $t$ is the time until merger at $t = 0$ \citep{Hendriks2024, 2025ApJ...990..211S}. 
This displacement along the line-of-sight relative to the observer equates to a 
delay in the GW arrival time given by $\tau = l/c$, where $c$ is the speed of light.
In our case, where a BBH is accelerated by a nearby third object BH of mass $m_3$, 
the maximum displacement in time is 
\begin{equation}
\tau(t) = \frac{l}{c} = \frac{1}{2} \frac{Gm_3}{c}\frac{t^2}{R^2}
\label{eq:tau}
\end{equation}
where $G$ is the gravitational constant, and $R$ is the distance to the third body from the BBH's
center-of-mass (COM). This displacement in arrival time, often referred to as the Romer-Delay,
can be translated into a GW phase shift. 
We define this phase shift as the observed angular displacement in units of radians between the reference
and the accelerated binary, approximately given by
\begin{align}
	\Delta{\phi}(t) & \approx 2\pi {\tau(t)}/{T(t)}, \nonumber\\
	 	    & \approx \frac{1}{2}\frac{G^{3/2}}{c} \frac{m_3 m_{12}^{1/2}}{R^2} \times \frac{t^2}{a(t)^{3/2}}
    \label{eq:dphi_general}
\end{align}
where we have used Eq.~\eqref{eq:tau}, $m_{12} = m_1 + m_2$ is the total mass of the BBH, 
and that the (inner) orbital period of the BBH at time $t$ can be estimated as,
\begin{equation}
T(t) = 2\pi\sqrt{a(t)^3/Gm_{12}}.
\end{equation}
Using the relations from \citet{Peters64}, $\Delta{\phi}(t)$ from Eq.~\eqref{eq:dphi_general} can be expressed in the circular limit ($e = 0$) as a function of time as,
\begin{equation}
\Delta{\phi}(t) \approx \frac{c^{7/8}G^{3/8}}{2} \left(\frac{5}{256}\right)^{3/8} \times \frac{m_3}{R^2}\frac{m_{12}^{1/8}}{m_1^{3/8}m_2^{3/8}} \times t^{13/8}.
\label{eq:dphi_e0_t}
\end{equation} 
This relation can be further expressed in terms of the evolving GW frequency, which in the circular limit equals $f = 2/T$, leading to 
\begin{equation}
\Delta{\phi}(f,e=0) \approx \frac{c^{9}G^{-7/3}}{2\pi^{13/3}} \left(\frac{5}{256}\right)^{2} \times \frac{m_3}{R^2}\frac{m_{12}^{2/3}}{m_1^{2}m_2^{2}} \times f^{-13/3}.
\label{eq:dphi_f_c}
\end{equation}
These equations are valid only for circular binaries; however, BBHs formed dynamically often retain eccentricities when observed. 
One way of defining the
GW phase shift in the eccentric limit is to keep the definition from 
Eq.~\eqref{eq:dphi_general}, 
with the addition that the evolution of the SMA $a$ then also depends on the time-evolving eccentricity.
Using the relation $a(e)$ from \citet{Peters64} in the high eccentricity limit, 
\begin{equation}
a(e) \approx \frac{2r_0e^{12/19}}{(1-e^2)}\frac{g(e)}{g(1)},\ (e_0 \approx 1).
\label{eq:ae_e1lim}
\end{equation}
where
\begin{equation}
g(e) = \left(1+121e^2/304\right)^{870/2299},
\end{equation}
and $r_0  = a_0(1-e_0)$, we can write Eq.~\eqref{eq:dphi_general} 
as,
\begin{align}
	\Delta{\phi}(e) & \approx \frac{288\sqrt{2}}{85^{2}g(1)^{13/2}} \frac{c^{9}}{G^{9/2}} \times \frac{m_3}{R^2}\frac{r_0^{13/2}}{m_1^{2}m_2^{2}m_{12}^{3/2}} \nonumber\\
	 	    &  \times e^{78/19}(1-e^2)^{1/2}g(e)^{13/2} \nonumber\\
		    & \approx \frac{288\sqrt{2}\pi^{-13/3}}{85^{2}g(1)^{13/2}} \frac{c^{9}}{G^{7/3}} \times \frac{m_3}{R^2}\frac{m_{12}^{2/3}}{m_1^{2}m_2^{2}} \times f_{0}^{-13/3} \nonumber\\
	 	    &  \times e^{78/19}(1-e^2)^{1/2}g(e)^{13/2},
    \label{eq:dphi_e_e1lim}
\end{align}
where $f_0$ is the GW peak frequency, $f$, and 
\begin{equation}
f(a,e) \approx \frac{1}{\pi}\sqrt{\frac{Gm_{12}}{r(a,e)^{3}}},
\label{eq:fp_rp}
\end{equation}
is evaluated at the initial values for $a$, $e$, i.e. $f_0 = f(a_0,e_0)$.
As seen in Eq.~\eqref{eq:dphi_e_e1lim}, the dependence on $e$ factors out into the
function $F(e) = e^{78/19}(1-e^2)^{1/2}g(e)^{13/2}$, with a maximum given by $e_m \approx 0.95$.
Using the approximate relation for the GW peak frequency given by Eq.~\eqref{eq:fp_rp},
Eq.~\eqref{eq:dphi_e_e1lim} can be approximated as \citep{2025ApJ...990..211S},
\begin{equation}
\Delta{\phi}(f) \approx \Delta{\phi}(f,e=0) \times (1+({f_0}/{f}))^7(1-({f_0}/{f}))^{1/2}.
\label{eq:f_f_app}
\end{equation}
Note here that when $f \gg f_0$ it transitions correctly into the circular case.

Finally, we note that our general definition of the GW phase shift from Eq.~\eqref{eq:dphi_general} is
natural and correct when the binary is circular, as it essentially provides a measure for how
well we can localize a possible displacement $\tau$ relative to the period $T$. However, when the
binary is eccentric, the period over which most of the GW signal is emitted is much shorter than the
orbital time, allowing for a much better resolution for measuring a possible
displacement, $\tau$ \cite[e.g.][]{2025arXiv251104540Z, 2025PhRvD.112f3005Z, 2025CQGra..42u5006T, 2025ApJ...991..131Z}.
If we denote the GW burst-width in the high eccentricity limit by $dt$, it follows that
\begin{equation}
dt \approx \frac{1}{f}
\end{equation}
where $f$ refers to the GW peak frequency from Eq.~\eqref{eq:fp_rp}. Using this instead of $T$ in
Eq.~\eqref{eq:dphi_general}, it follows that
\begin{align}
	\Delta{\phi}_p & \approx 2\pi {\tau(t)}/{dt}, \nonumber\\
	 	    & \approx (n_p/2) \times \Delta{\phi}
    \label{eq:dphi_nharm}
\end{align}
where $n_p \geq 2$ is the $n$'th-harmonic corresponding to the GW peak frequency harmonic,
and $\Delta{\phi}$ is the GW phase shift defined in Eq.~\eqref{eq:dphi_general}.
The value for $n_p$ will be $\simeq 2(1+e)^{1.1954}/(1-e^2)^{3/2}\gg 2$ 
in the high eccentricity limit \citep{Wen2003,Hamers2021}, indicating that the GW phase shift can
therefore be significantly larger than the one directly carried over from the circular limit.
In this study, we primarily consider results from the conservative estimate based on the
fundamental harmonic with the GW phase shift given by Eq.~\eqref{eq:dphi_general}, but we will also
show a few cases where the definition from Eq.~\eqref{eq:dphi_nharm} is used. As waveforms required for
a full signal-to-noise (S/N) analysis are not yet available, we reserve any S/N calculations for upcoming
studies.

\subsection{AGN Binary Black Hole Merger Types}

BBHs merging through different dynamical path ways will be considered differently in our analysis, as they give rise to different outcomes and observables.
In the following, we outline and define the three main merger channels arising from our AGN disk models. 

{\bf (1) Binary Merger (non-GWC):} This type of merger happens between two BHs that
evolve in the AGN disk while not being bound to anything else than each other and the
central SMBH. Such BBHs are often driven together through a combination of
gas-drag and GW emission \citep{Bartos17,Tagawa19}. They might have undergone previous
interactions with other objects while migrating through the disk, but none of these objects
are near at merger, which further implies that they generally merge on near
circular orbits. For calculating the associated GW phase shift, we use the
equations from Sec. \ref{Gravitational Wave Phase Shifts} with $m_3$ and $R$ set to
the mass of the SMBH and the distance from the BBH COM to the SMBH, respectively.
The orbital elements $a$ and $e$ for each BBH merger is taken directly from the output of our AGN model. Throughout the paper we refer to these mergers as {`non-GWC'}.

{\bf (2) Single-Single Merger (GWC-SS):} These mergers are formed through a
GW single-single capture between two initially unbound BHs. In our models,
this process is often taking place in the inner parts of the AGN disk.
For calculating the GW phase shift of these mergers we follow the same procedure as for the `non-GWC' mergers above. Throughout the paper we refer to these as {`GWC-SS'}.

{\bf (3) Binary-Single Merger (GWC-BS):} These mergers occur during a
binary-single interaction, while the third object is still bound to the
merging BBH (see \citealt{Samsing14,Samsing17, Samsing18b}).
This type of merger is particularly
interesting, as the third object often is so close that the
resultant acceleration on the BBH COM dominates over the  acceleration from the SMBH. 
These types of mergers can therefore result in relatively high GW phase shifts.
In addition, if one takes into account that the surrounding gas 
can harden the system while the three BHs are interacting \citep{Rowan2025},
the GW phase shift can get even larger. We explore both cases in the following.
Without gas effects we calculate the GW phase shift by setting $m_3$ equal to the third bound
BH, and $R$ equal to the initial SMA of the BBH before the interaction as this
sets the characteristic length scale of the system. When gas is included, we scale
down the scales of the entire system by a fraction inspired from full
hydro-simulations of triple scatterings in an AGN disk \citep{Rowan2025}.
We acknowledge these prescriptions are only approximate, but it is beyond this work to
perform a full PN/gas few-body scattering exploration for all interactions
that might take place in our AGN disk models (see \citealt{Tagawa20_MassGap}).
Throughout the paper we refer to these mergers as {`GWC-BS'}.

The majority of merging binaries ($\sim90\%$) were initially formed through the gas-capture mechanism \citep{Tagawa19}, where a fraction of the energy of the encounter is
dissipated via the gas \citep[see][]{Rowan2023,Rowan2024_II,Whitehead2023,Whitehead2025}.
The gas-capture model in our four models was compared with that derived from hydrodynamical simulations in \cite{Rowan2024_rates}, showing good agreement.

\section{Results}\label{Sec:Results}

In the following sections, we present the distribution of the phase shift expected
in the AGN channel, and discuss the effects of gas and eccentricity on the phase shift.

\subsection{Distribution of Gravitational Wave Phase Shifts}

\begin{figure}
    \centering
    \includegraphics[width=1\columnwidth]{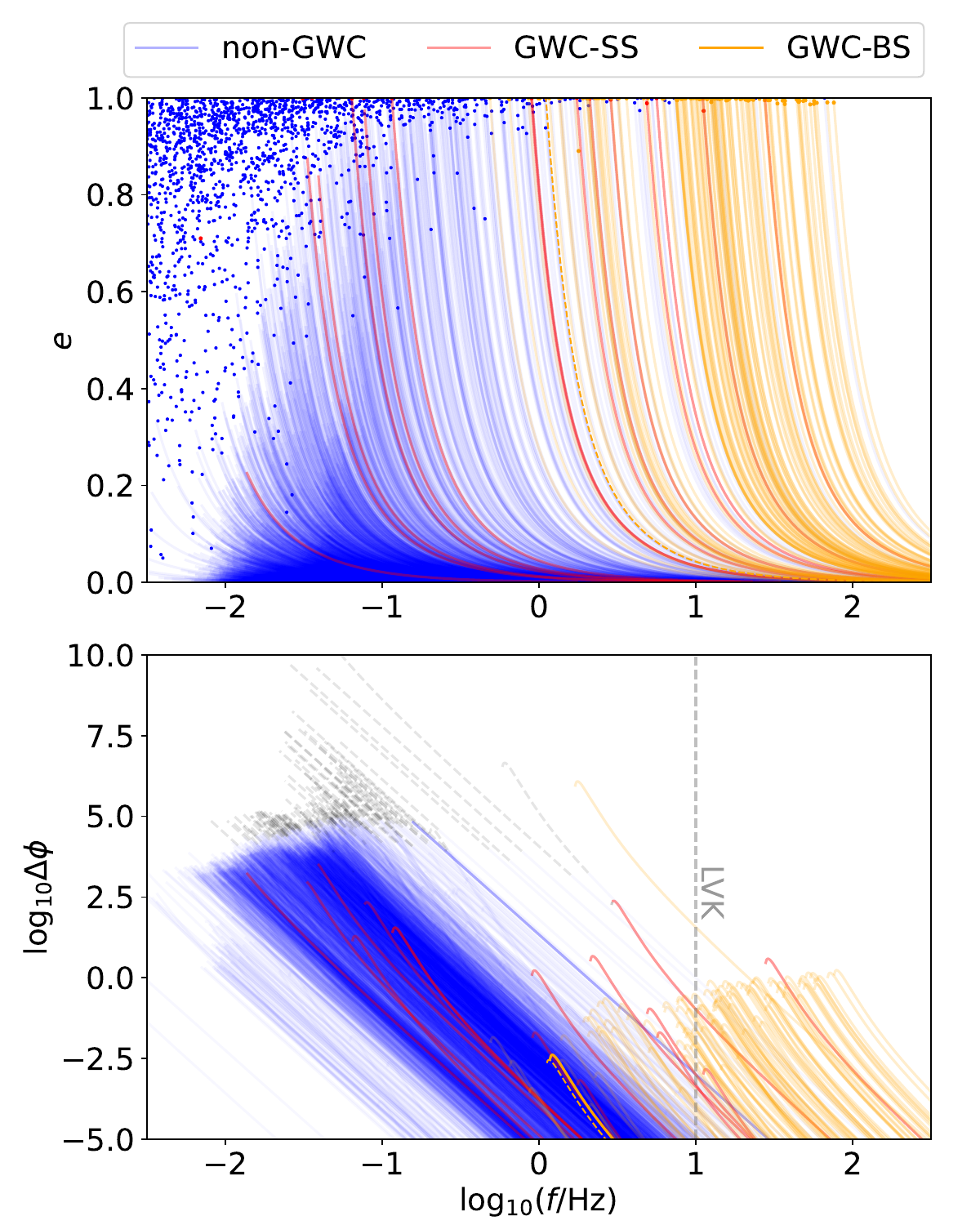}
    \caption{
    The evolution of the orbital eccentricity and the cumulative phase shift until mergers for model~{\bf M1}. 
    Points represent values at binary formation, 
    and lines represent their evolution. 
    Blue, orange and brown, and red lines display 
    the results for mergers due to non-GWC processes, mergers due to the GWC mechanism during
    binary-single interactions, and those during single-single interactions, respectively.     
    For GWC binaries during binary-single interactions, the solid lines correspond to cases
    where the acceleration from the bound third BH is dominant,
    while the dashed lines correspond to cases where the acceleration from the SMBH is dominant. 
    In the lower panel, the dashed gray lines depict the evolution for phases in which the binary moves more than $1/4$ of the orbit around its companion (third BH or SMBH) before merging. 
    The lines are shown for the final $5~{\rm yr}$ to merger, 
    and are not connected to the points in the upper panel for cases with long merger time. 
    The gray vertical dashed line in the lower panel represents the typical value for
    the LVK sensitivity band.   
    }
    \label{fig:e_phi_m1}
\end{figure}

\begin{figure}
    \centering
    \includegraphics[width=1\linewidth]{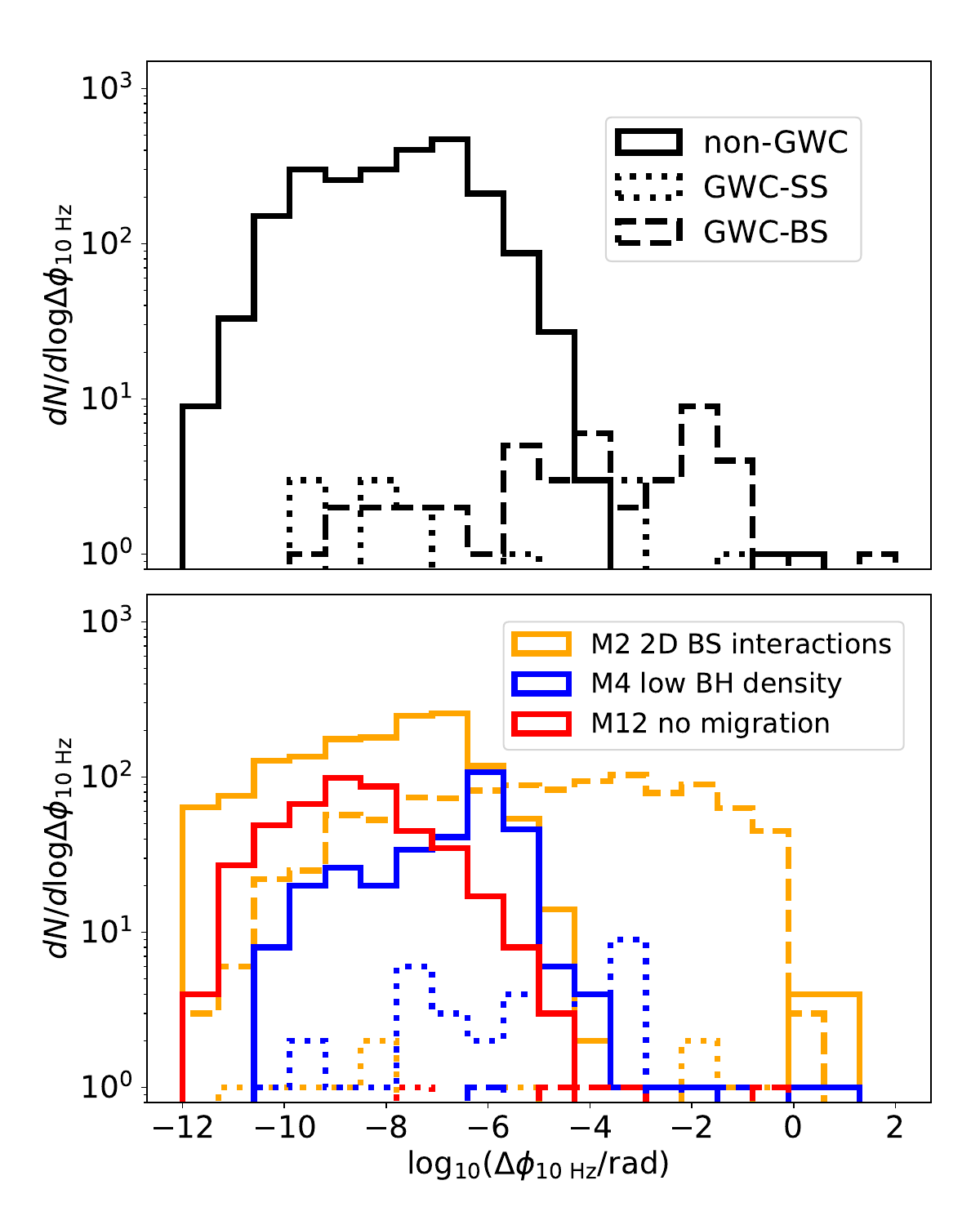}
    \caption{
    The distribution of 
    cumulative phase shifts above $10~{\rm Hz}$.   
    Solid, dashed, and dotted lines represent 
    the results for mergers due to non-GWC processes, mergers due to the GWC mechanism during
    binary-single interactions, and those during single-single interactions, respectively. 
    The upper panel shows the result for model {\bf M1}, and orange, blue, and red lines in 
    the lower panel shows those for models~{\bf M2}, {\bf M4}, and {\bf M12}, respectively.}
    \label{fig:phi_dist}
\end{figure}

\begin{figure}
    \centering
    \includegraphics[width=1\linewidth]{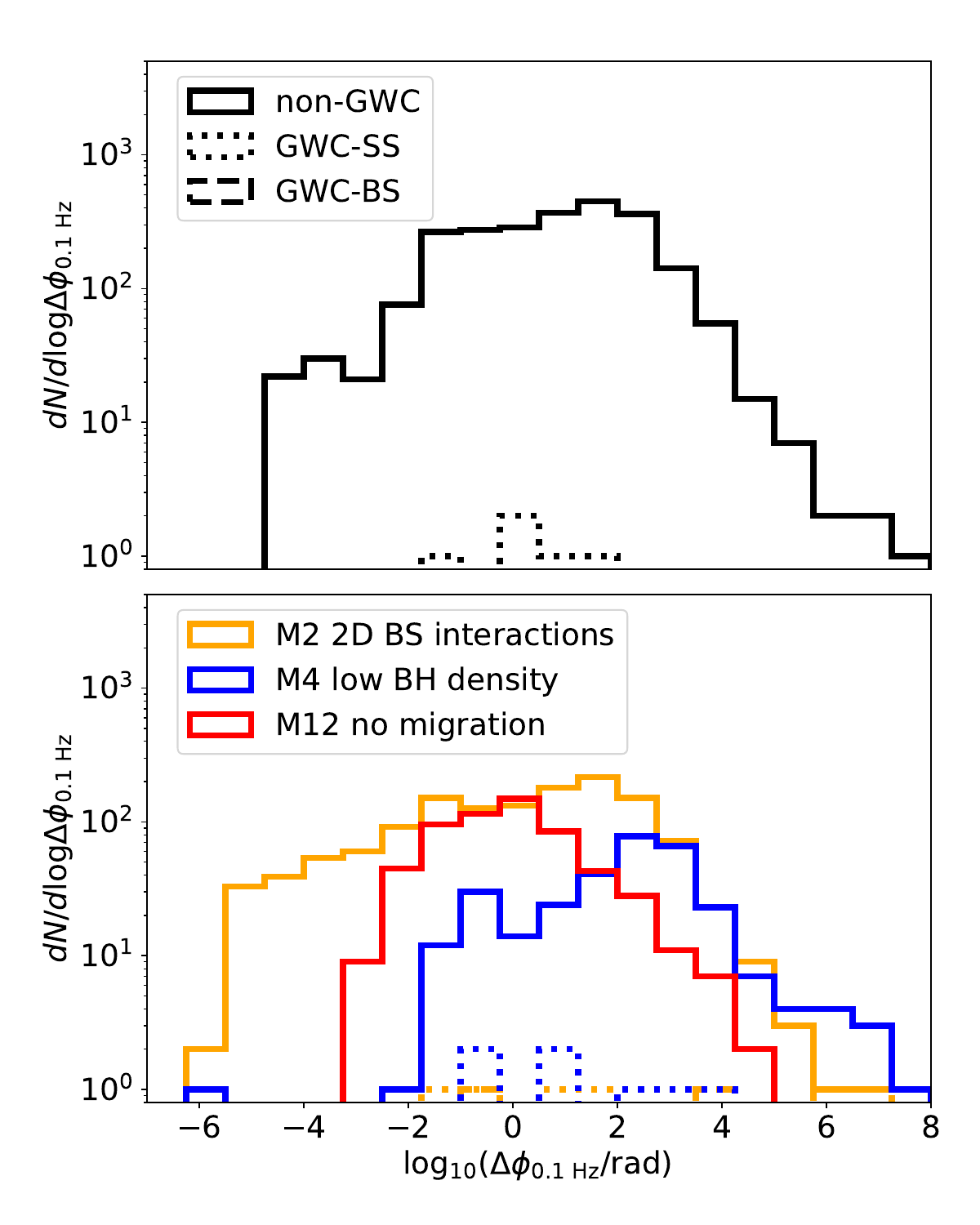}
    \caption{
    Same as in Fig.~\ref{fig:phi_dist}, but 
    for cumulative phase shifts above $0.1~{\rm Hz}$. 
    }
    \label{fig:phi01_dist}
\end{figure}

Results for our fiducial {\bf M1} model are presented in Fig.~\ref{fig:e_phi_m1},
which shows the evolution of the eccentricity (top) and the GW phase shift (bottom),
respectively, as a function of the peak GW frequency, for our considered merger types.
As seen in the figure, binary mergers (non-GWC) primarily form at low GW peak
frequencies of $\lesssim 0.01$--$0.1~{\rm Hz}$, with an eccentricity distribution
shaped by a combination of previous scatterings and the mechanisms behind the binary
assembly. This population does exhibit a significant GW phase shift caused
by the acceleration of the central SMBH, but only at lower frequencies as it
rapidly decreases $\propto f^{-13/3}$ \citep{Zwick2025}. This type of GW phase shift
has been studied in the literature before \citep{Vijaykumar2023}, and a recent study even presents tentative evidence of acceleration in a similar configuration \citep{Yang2025}.
The majority of the non-GWC population will be a prime observational target for
upcoming deci-Hertz detectors such as DECIGO \citep{Yagi2011}, DO \citep{ArcaSedda2020},
and ALIA \citep{Mueller2019}. Distributions of the cumulative phase shifts above
$10~{\rm Hz}$ and $0.1~{\rm Hz}$ are presented in
Figs.~\ref{fig:phi_dist} and \ref{fig:phi01_dist}, respectively. 

GWC-BS mergers primarily form with an initial peak GW frequency of $\gtrsim 1~{\rm Hz}$,
as the third bound object imposes natural limits on the initial SMA
and eccentricity of the merging BBH \citep{Samsing14}. 
This influence of the third object also leads to GW phase shifts that are
significantly higher than those arising
from a distant SMBH. The GWC mergers therefore constitute a distinct population, that
we observe in our AGN disk models, typically exhibiting phase shifts of
$\gtrsim 0.1~{\rm rad}$ near the LVK band and even beyond. GWC-SS mergers form differently,
and their acceleration are entirely due to the SMBH and how close the disk model allows
them to form. Therefore, these mergers can occur at all frequencies with large phase
shifts, as also seen in the figure. While the GWC-BS mergers can be studied with
local scattering experiments, the evolution of GW-SS mergers and their phase shifts
depend sensitively on all the binary evolution mechanisms through the disk and the
global disk properties.

\subsection{Distribution of Binary Accelerations}

\begin{figure}
    \centering
    \includegraphics[width=1\columnwidth]{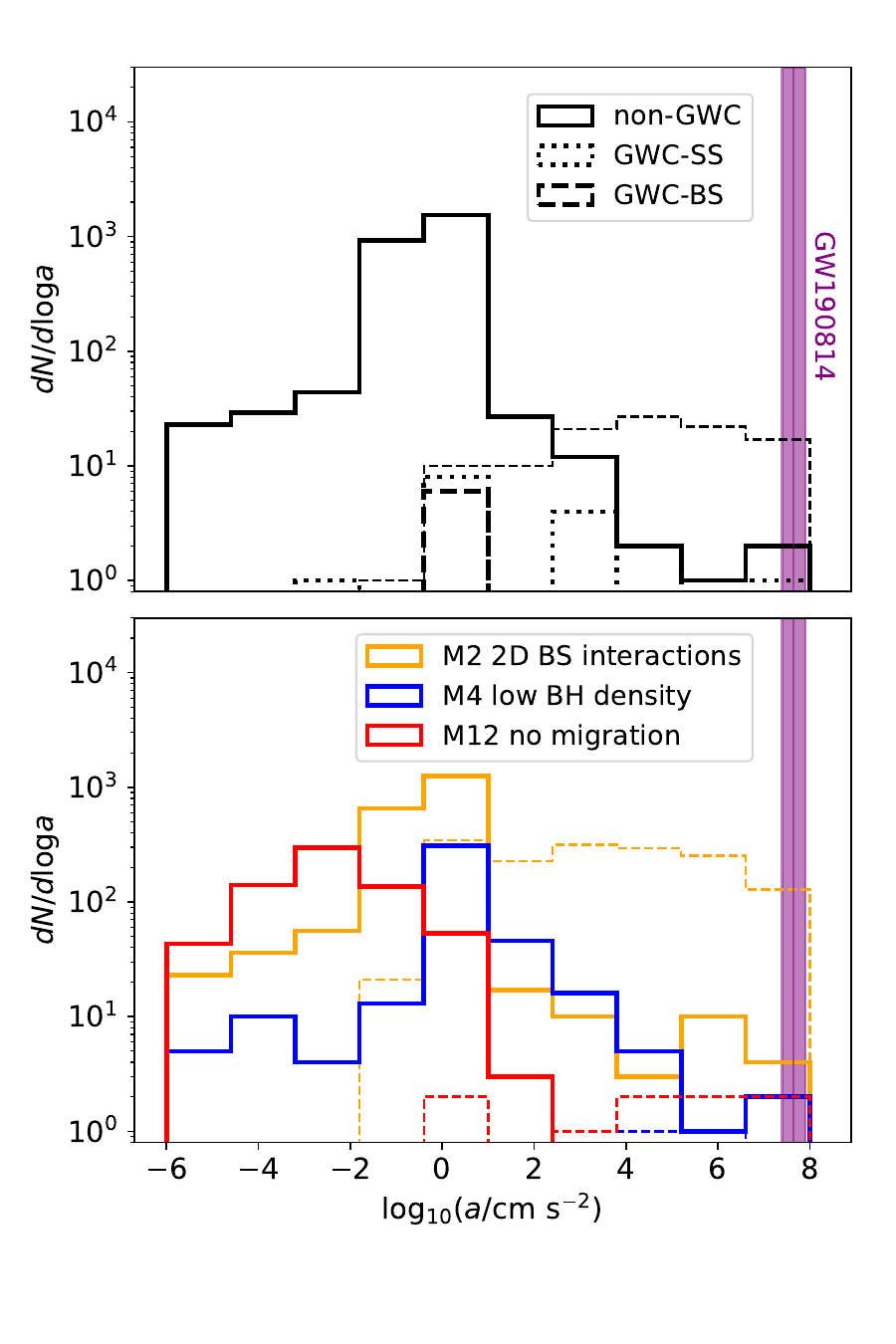}
    \caption{ 
    The distribution of acceleration from the third body at binary formation. 
    In the upper panel, solid, dotted, and dashed lines represent 
    the results for model~{\bf M1} for mergers occurring due to non-GWC processes, mergers resulting
    from the GWC mechanism during binary-single interactions, and those occurring during single-single
    interactions, respectively. In the lower panel, blue, orange, and red lines represent 
    the results for all the mergers for models~{\bf M2}, {\bf M4}, and {\bf M12}, respectively. 
    In upper and bottom panels, thick lines indicate 
    cases where the acceleration from the SMBH is dominant, 
    while thin dashed lines indicate 
    cases where the acceleration from the third bound BH is dominant.    
    The purple vertical box and line represent 90$\%$ confidence intervals and median for acceleration in GW190814 suggested by \citet{Yang2025}.
    }
    \label{fig:a_dist}
\end{figure}

Fig.~\ref{fig:a_dist} shows the distribution of the BBH COM acceleration for our
considered merger types. The top plot shows results for the {\bf M1}-model, where
the bottom plot shows the total distribution across all three merger types for the
remaining models, {\bf M2}, {\bf M4}, and {\bf M12}. The solid- and dashed lines illustrate the acceleration
on the BBH COM dominated by the SMBH and a nearby third-object, respectively. As seen,
non-GWC mergers peak at around $\sim 1~{\rm cm~s^{-2}}$ and GWC mergers peak at much
higher values. Notably, these accelerations are significantly larger than those typically expected for the triple channel \citep{Antonini2017,Antonini2018}. 
To distinguish the underlying environments, detecting these accelerations is crucial.

According to \citet{Zwick2025}, for BH masses ranging from $5$ to $200~\Msun$ and
redshifts between $0$ and $5$, a significant fraction of mergers with accelerations
of $a\sim 10^2$--$10^4~{\rm cm~s^{-2}}$, $a\gtrsim 10^6~{\rm cm~s^{-2}}$, and
$a\gtrsim 10^7~{\rm cm~s^{-2}}$ are detectable by the Einstein Telescope and
Cosmic Explorer, as well as during the $A\#$ and O4 sensitivity observations of
LIGO/Virgo/KAGRA, respectively. As seen in Fig.~\ref{fig:a_dist}, a certain fraction of
mergers meets these conditions and may therefore serve as promising signatures for
mergers in AGN disks, potentially detectable even by the O4 sensitivity.
These large accelerations are also consistent with the
value ($\sim 0.0015\mathrm{c}\,{\rm s}^{-1}$) implied for GW190814  \citep{Yang2025}
shown in purple in Fig.~\ref{fig:a_dist}, which could suggest a GWC-BS origin.
If binary-single interactions occur in a two-dimensional geometry (model~M2),
a larger fraction of the GW mergers exhibit higher GW phase shifts, as seen in the lower
panel of Fig.~\ref{fig:phi_dist}, due to the significantly higher occurrence rate
of GWC-BS mergers \citep{Samsing20}. Additionally, the fraction of mergers with
significant phase shifts ($\gtrsim 10^{-2}~{\rm rad}$) is less affected by the
number density of BHs, although the total number of mergers is influenced (blue lines). 
Similar to the fiducial model, binaries undergo multiple binary-single interactions
before mergers, resulting in a comparable fraction of GWC mergers. 
The same reasoning applies to the model without migration (red lines). 
The fraction of mergers with a cumulative phase shift exceeding $0.1~{\rm rad}$ above
10 Hz is $0.2\%$, $1\%$, $0.4\%$, and $0.3\%$ for
models~{\bf M1}, {\bf M2}, {\bf M4}, and {\bf M12}, respectively, where the fractions
exceeding $1~{\rm rad}$ are $0.07\%$, $0.2\%$, $0.4\%$, and $<0.2\%$.

\begin{figure}
    \centering
    \includegraphics[width=1\columnwidth]{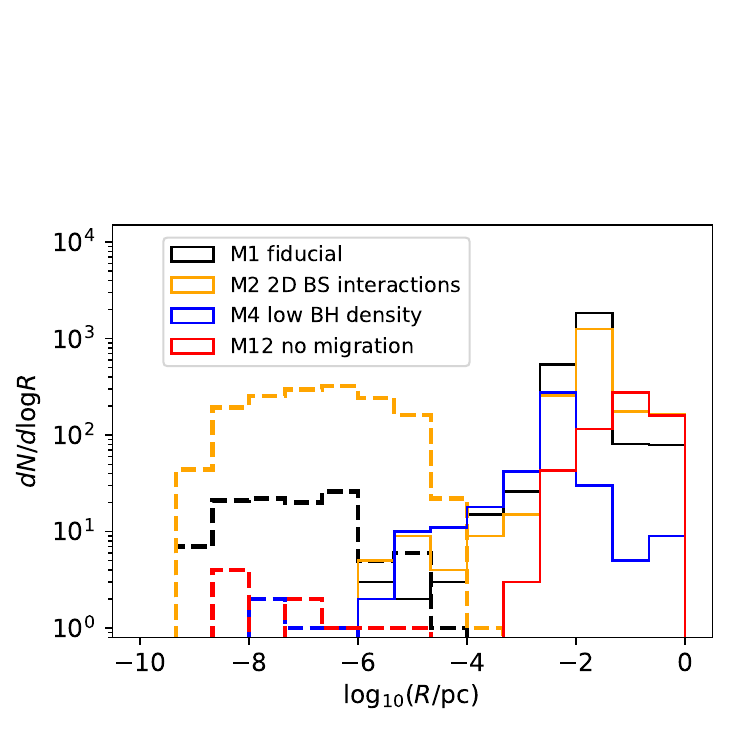}
    \caption{
    The distribution of the distance to the third body.    
    Black, orange, blue, and red lines represent the results for models~M1, {\bf M2}, {\bf M4}, and {\bf M12} respectively.     
    The thick lines represent the distance to the SMBH in cases where the acceleration from the SMBH is dominant, 
    while dashed lines represent the distance to the third bound BH in cases where the acceleration from the BH is dominant. 
    }
    \label{fig:r_dist}
\end{figure}

Given a fixed SMBH mass, the amount of acceleration is determined by the distance
to the SMBH, $R$, whose distribution is represented by the solid lines
in Fig.~\ref{fig:r_dist} for all four models. In the fiducial
{\bf M1}-model (black lines), distances are concentrated around $\sim 0.01~{\rm pc}$, 
as migration slows down in these regions due to the formation of deep
gaps \citep{Tagawa19,Gilbaum2025}. In the {\bf M12}-model without migration
(solid red lines in Fig.~\ref{fig:r_dist}), 
mergers occur at $\sim 0.1~{\rm pc}$, resulting in lower acceleration from
the SMBH and smaller phase shifts (as shown by the solid red lines in the lower
panels of Figs.~\ref{fig:a_dist} and \ref{fig:phi01_dist}). 

Although we have not explicitly included migration traps or thermal torques in
the fiducial model, we consider their potential effects. Incorporating migration traps
into the fiducial model, following the methodology described by \citet{Bellovary16}, 
BH mergers do not occur at $R\leq 10^{-4}~{\rm pc}$. 
Consequently, both non-GWC and GWC-SS mergers with $a\gtrsim 10^4~{\rm cm~s^{-2}}$ and
$\Delta \Phi \gtrsim 10^{-3}~{\rm rad}$ at 10 Hz are suppressed. 
Conversely, by considering thermal torques in the prescriptions
described by \citet{Jimenez2017} and \citet{Grishin2024}, 
BHs are driven toward $R\leq 10^{-4}~{\rm pc}$ likely due to gap formation and
saturation effects. The resulting distributions for acceleration and phase shift are
similar to those in the fiducial model. Since mergers occurring at $R\leq 10^{-4}~{\rm pc}$
are relatively minor, the influence of these prescriptions on the phase shift remains limited.

\subsection{Gas effects}

The effects of gas during binary-single interactions are not considered or included in
any of the dynamical prescriptions used in the AGN models employed
here \citep{Tagawa20_ecc}. This omission is largely justified if gas drag is described
by the formulas for gas dynamical friction \citep{Ostriker99}, as this generally would
lead to very small effects \citep{Tagawa19} due to the relatively high velocities
close to merger. However, recently \citet{Rowan2025} performed two-dimensional
hydrodynamical simulations of binary-single scatterings in a disk-like environment and found
that the entire triple system typically undergoes significant shrinkage due to
interactions with the surrounding gas. Furthermore, since the 
kinetic energy of the interacting objects is generally rapidly dissipated, it was found
that the interactions can involve many more close encounters compared to the 
vacuum case before terminating (see also \citealt{WangMengye2025_BS1, WangMengye2025_BS2}). 
Shrinkage of the system is efficient only during interactions, presumably
because small Bondi-Hoyle-Lyttleton radii enable gas dynamical friction
from high-density gas nearby BHs. Specifically, \citet{Rowan2025} showed that
triple systems during the interaction can easily shrink by a factor of
$\sim 20$, with significant implications for the resultant phase shifts.

Here we construct a model for gas drag that better aligns with the simulation results. 
According to \citet{Rowan2025}, the captured gas accumulates around the binary system. 
Once binary-single interactions occur, this accumulated gas facilitates the binary evolution,
by dissipating the kinetic energy of the third body. Assuming that a significant fraction of the captured
gas is ejected at an orbital speed of the binary ($v_{\rm orb}$) multiplied by a factor ($f_{\rm ej}\sim \mathcal{O}(1)$), 
we tentatively prescribe the energy dissipation rate as  
\begin{eqnarray}
\label{eq:dedt}
\frac{d{E}}{d{t}}\sim \frac{{M}_{\rm ret}f_{\rm ej}^2 v_{\rm orb}^2}{t_{\rm decay}},
\end{eqnarray}
where $M_{\rm ret}$ is the retained gas mass, 
defined as the accumulated captured gas mass minus the depleted gas mass, and $t_{\rm decay}$ is the timescale of binary
evolution during binary-single interactions, typically $\sim 1$--$100$ times the dynamical time of the binary \citep{Samsing14,Rowan2025}. 
The uncertainty in this timescale does not affect the results below. 
Using $v_{\rm orb}^2\sim G(m_1+m_2)/R$ and $E=Gm_3(m_1+m_2)/R$, Eq.~\eqref{eq:dedt} can be reformulated as 
\begin{eqnarray}
\label{eq:e_e0}
E\sim E_0 {\rm exp} \left( \frac{f_{\rm ej}^2 {M}_{\rm ret} t}{m_3 t_{\rm decay}}\right),
\end{eqnarray}
which is valid for $t\leq t_{\rm decay}$. 
Using the results from the simulations of \citet{Rowan2025}, 
and evaluating the gas capture rate with Eq.~(1) of \citet{Tagawa2022_BHFeedback} 
(which can predict capture rates, at least for two-dimensional simulations, \citealt{Whitehead2023}), 
we find $f_{\rm ej}\sim 1.1$ (although the variability is high, see Fig.~8 of \citealt{Rowan2025}). 
This suggests that the assumption that most captured gas is ejected at a velocity $\sim v_{\rm orb}$ is consistent
with the simulation. Note that in \citet{WangMengye2025_BS2}, gas disappears due to accretion and $M_{\rm ret}$ is somewhat uncertain, making this evaluation more challenging.

Since three-body systems shrink due to gas effects, we need to consider the resulting changes in the properties
of GWC binaries at formation. When the distance to the third body is reduced by some factor, the
pericenter distance within which GWC binaries form also changes. 
The maximum pericenter distance is proportional to the relative velocity as
$r_{\rm p,max} \propto v^{-4/7}$ \citep[e.g.][]{OLeary09}. 
Assuming that the relative velocity is roughly given by the Keplerian velocity of the binary and
that the Keplerian velocity scales as ($v\propto R^{-1/2}$), then the maximum pericenter distance
then scales as $r_{\rm p,max}\propto R^{2/7}$ \citep{Samsing14, Samsing18b, 2021ApJ...923..126S}. We assume that the initial semi-major axis of the
binary is reduced by the same factor as the reduction in $R$ caused by gas effects. 
The eccentricity at GWC binary formation is adjusted to ensure that the pericenter distance decreases
by the same factor corresponding to the reduction of $r_{\rm p,max}$. 
In this way, both the initial peak frequency of the binary and the cumulative phase shift are affected,
reflecting the shrinkage of the three-body systems due to gas effects. 

In realistic environments, we can expect the retained mass to saturate because of depletion through
accretion and wind mass loss. In our simulation, assuming that the captured gas can be retained around the
binary for the viscous timescale, we set $M_{\rm ret}\sim {\dot M}_{\rm cap}t_{\rm vis}$, where 
${\dot M}_{\rm cap}$ is the gas capture rate, and $t_{\rm vis}$ is the viscous timescale at the
circularization radius for the captured gas. The circularization radius is set
to $r_{\rm circ}=f_{\rm circ} r_{\rm Hill}$, where $r_{\rm Hill}$ is the Hill radius, and
$f_{\rm circ}$ is a factor representing the ratio between the circularization radius and the
Hill radius \citep{Sagynbayeva2025}. Since the capture rate is highly super-Eddington, 
we assume the disk is thick, with $t_{\rm vis}\sim t_{\rm orb}(r_{\rm circ})/\alpha_{\rm ss}$,
where $t_{\rm orb}(r)$ is the orbital timescale at radius $r$ from the binary center,
and $\alpha_{\rm ss}\sim 0.1$ is the alpha-viscosity parameter. 
By incorporating $t=t_{\rm decay}$ and $f_{\rm ej}=1.1$ into Eq.~\eqref{eq:e_e0}, we can calculate
the final energy of the three-body system. 

Conversely, when $M_{\rm ret}\gg m_3$, Eq.~\eqref{eq:e_e0}
overestimates the final energy because, before all the retained gas is ejected, the binary can
merge via the GWC mechanism. \citet{Rowan2025} estimated that during $\sim 20$ intermediate states in
binary-single interactions, once the binary is hardened to $\lesssim 10^4$ Schwarzschild
radii, $\gtrsim 80\%$ of binaries merge. Reflecting this, we assume that binaries are maximally
hardened to $\sim 10^4$ times the Schwarzschild radius of the primary BH.

\begin{figure}
    \centering
    \includegraphics[width=1\linewidth]{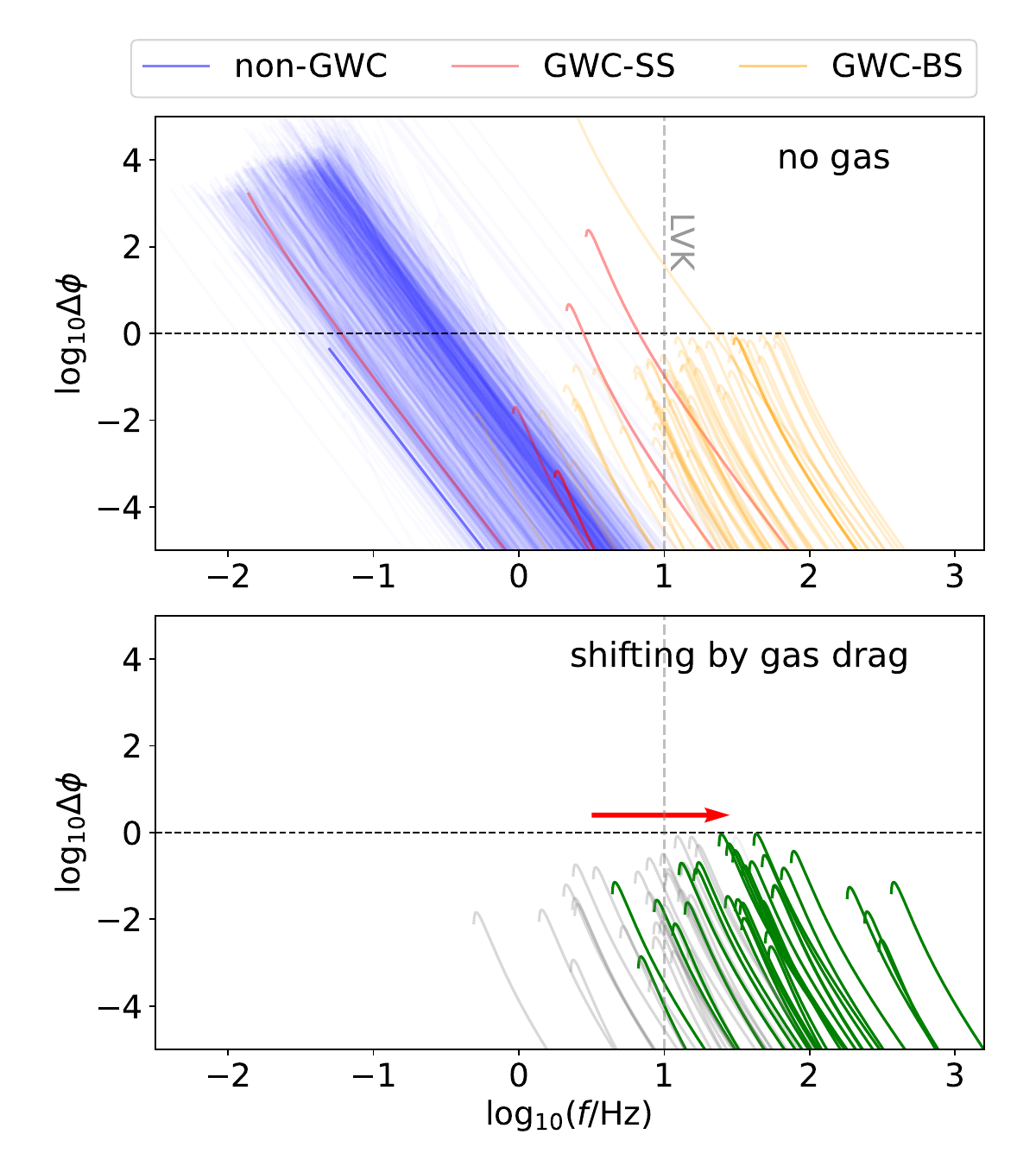}
    \caption{
    The upper panel is the same as the lower panel of Fig.~\ref{fig:e_phi_m1}. 
    In the lower panel, 
    the results for GWC-BS mergers are shown. 
    The cases in which the distance to the third bound body is reduced by a factor of $\geq 1.1$ due to gas effects are represented by 
    green lines, while those without gas effects are shown with gray lines. 
    Cases where the distance change is $\leq 1.1$ are not displayed.
    }
    \label{fig:phi_f_gas}
\end{figure}

Fig.~\ref{fig:phi_f_gas} compares the cumulative phase shifts without (upper panel) and with gas (lower panel). 
The distance to the third body is reduced by gas effects by a factor
of $\gtrsim 1.1$ in $\sim 30\%$ of the GWC mergers during binary-single interactions (green lines).
These are more frequent at lower-frequency bands, since binaries at higher frequencies often exceed
the limit of $\sim 10^4$ times the Schwarzschild radius. When the distance to the third body is reduced
by the gas effect, the GW frequency is shifted to higher values, while the cumulative phase shift is less affected. 
This is because the GW frequency scales as $f_0\propto r_0^{-3/2}\propto R^{-3/7}$, while 
the cumulative phase shift scales as $\propto f_0^{-13/3}R^{-2}\propto R^{-1/7}$ (Eq.~\ref{eq:dphi_e_e1lim}). 
Thus, the dependence of the cumulative phase shift on
$R$ is weak \citep[e.g.][]{2025ApJ...990..211S}. As the GW frequency typically shifts
above $\sim 10~{\rm Hz}$, the phase shift may be more easily constrained by LIGO/Virgo/KAGRA, 
since it accumulates predominantly in frequency bands where these detectors are more sensitive. 
Importantly, these results suggest that a significant fraction of non-GWC mergers may also merge as GWC mergers due to the boosting of the peak frequency by gas effects. This would have a substantial impact on observables and 
is worth assessing in future studies.

Additionally, we have not included the effects of gas on the GWC mechanism during single-single
encounters or for non-GWC mergers. In the former case, the effective masses of the binary components
are enhanced by surrounding circum-BH disks \citep{Rowan2024_II}, which increases the probability of
the GWC mechanism, as the maximum pericenter distance within which GWC can occur is proportional to
the total mass of the binary, assuming a fixed mass ratio \citep{Quinlan1989}. 
Incorporating this effect could further boost the rates of mergers with high eccentricities
and significant phase shifts. 

Furthermore, gas can cause a phase shift through its torques. This effect would be more easily observable
for non-GWC mergers, since the time from the last scattering to merger is longer. 
Such torques are expected to be detectable by LISA, TianQin, and Taiji \citep{DuttaRoy2025,ChenRan2025}.

\subsection{Eccentric Burst Sources}

\begin{figure}
    \centering
    \includegraphics[width=1\linewidth]{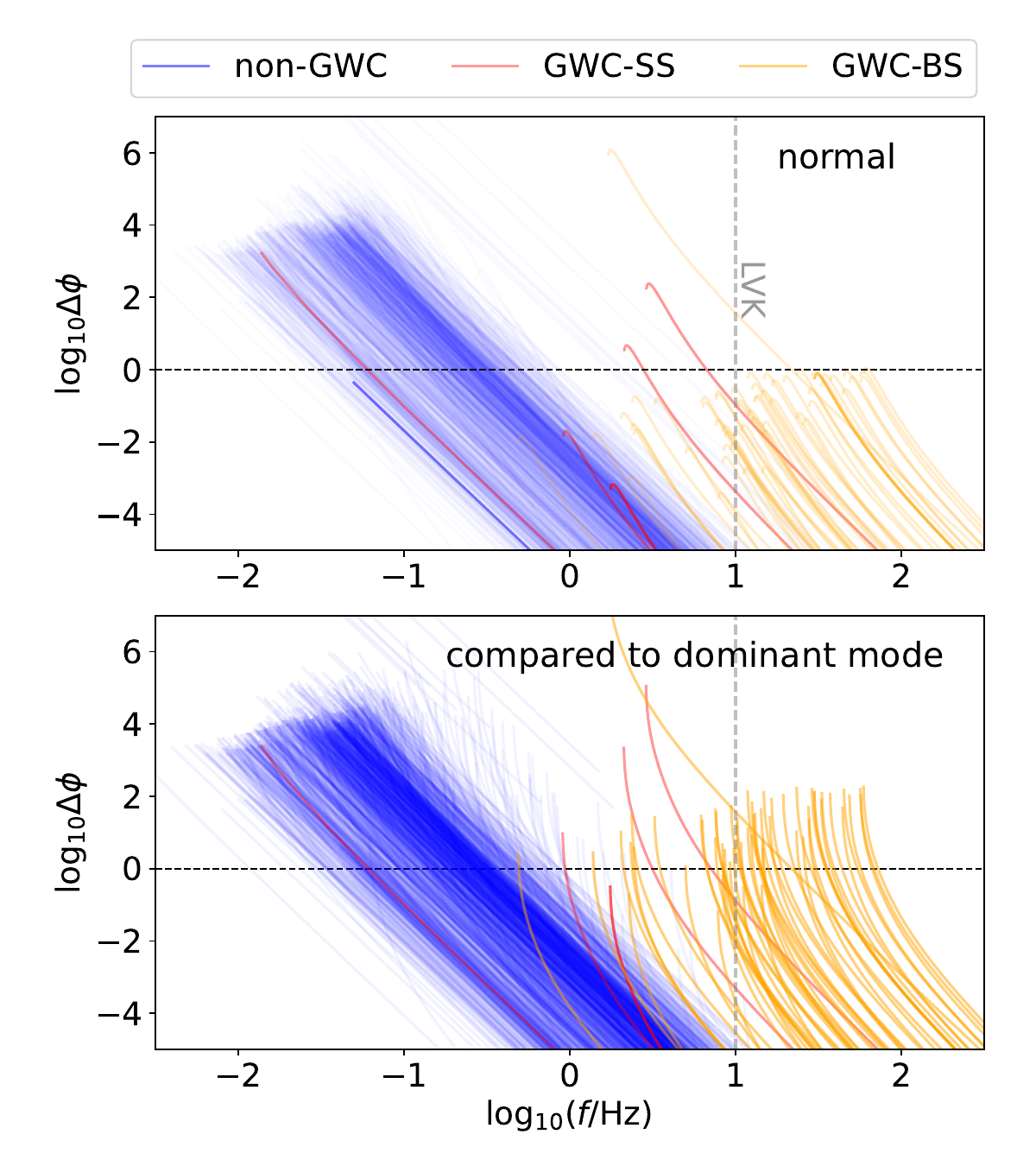}
    \caption{
    The upper panel is the same as the lower panel of Fig.~\ref{fig:e_phi_m1}. 
    In the lower panel, 
    we consider the phase shift relative to the period of the GW mode that predominantly contributes to the amplitude of the GW emission, as given by Eq.~\eqref{eq:dphi_nharm}.
    }
    \label{fig:phi_fp}
\end{figure}

We have defined the phase shift as the change in the number of cycle counts of the GW signal, with
the frequency given by the orbital frequency. 
This limit naturally bridges to the well studied circular case.
However, when the binary becomes highly eccentric, higher modes dominate the GW signal, which in fact allows
for a much better determination of the phase shift as also argued in Sec. \ref{Gravitational Wave Phase Shifts} and
recently in \citet{2025arXiv251104540Z, 2025PhRvD.112f3005Z, 2025ApJ...991..131Z} and \citet{2025CQGra..42u5006T}. To illustrate how the phase shift is enhanced by considering different modes, 
Fig.~\ref{fig:phi_fp} presents the phase shift defined by the change in cycles calculated with the
orbital frequency (upper panel) and that with the dominant GW frequency (lower panel).  

For binaries formed via GWC (orange and red lines in Fig.~\ref{fig:phi_fp}), 
the high eccentricities cause the phase shift relative to cycles with the dominant mode frequency
to exceed $\sim 1~{\rm rad}$ after binary formation. Thus, the phase shift becomes significant depending on
whether it is measured with respect to the orbital frequency or the dominant GW mode frequency. 
Assessing the detectability of these shifts in highly eccentric systems is a topic for future work.

\section{Conclusions}
\label{sec:conclusions}

In this study, we have predicted the distribution of phase shifts caused by acceleration from nearby objects for BH mergers in AGN disks. 
These predictions are based on a semi-analytical model combined with one-dimensional $N$-body simulations. 
Our key findings are as follows:

\begin{enumerate}

\item 
For all mergers, binaries experience strong acceleration from the central SMBH. 
Non-GWC binaries undergo moderate acceleration of $\sim {\rm cm~s^{-2}}$, 
which is detectable by future GW facilities such as the Einstein Telescope and Cosmic Explorer. 

\item 
GWC binaries formed during binary-single interactions experience strong accelerations caused by a third body near the merging binaries, resulting in significant phase shifts. 
Additionally, GWC binaries formed during single-single interactions are located close to the SMBH. 
These binaries are subject to strong accelerations, which are larger than those expected in other formation channels. 
Furthermore, a small fraction ($\sim 0.1$--$1\%$) of mergers exhibit a substantial cumulative phase shift of $\gtrsim 0.1$--$1~{\rm rad}$ above $10~{\rm Hz}$, which could be detectable with current GW detectors. 

\item 
Gas effects may shift the GW frequency of non-GWC and GWC mergers formed during binary-single interactions, enhancing the prospects for detecting phase shifts by LIGO/Virgo/KAGRA. 
The influence of gas effects merits further elucidation through additional studies. 

\item 
The phase shift from highly eccentric mergers could be significant, due to the enhancement of the frequency in the dominant GW mode, as also newly pointed out by \citet{2025arXiv251104540Z,2025PhRvD.112f3005Z,2025ApJ...991..131Z} and \citet{2025CQGra..42u5006T}.

\end{enumerate}

\acknowledgments

Simulations were carried out on Cray XD2000 at the Center for Computational Astrophysics, National Astronomical Observatory of Japan. 
J.S. and K.H are supported by the Villum Fonden grant No. 29466, and by the ERC Starting Grant no. 101043143 – BlackHoleMergs.
L.Z. is supported by the European Union’s Horizon 2024
research and innovation program under the Marie Sklodowska-Curie grant agreement No. 101208914.
J.T. is supported by the Alexander von Humboldt Foundation under the project no. 1240213 - HFST-P.
The Center of Gravity is a Center of Excellence funded by the Danish National Research Foundation under grant No. 184.

\appendix

\bibliographystyle{aasjournal}
\bibliography{agn_bhm}

\end{document}